\documentclass[letterpaper, 10 pt, conference]{ieeeconf}  

\IEEEoverridecommandlockouts                              

\usepackage{graphicx} 
\usepackage{amssymb}  
\usepackage{amsmath,amsfonts}
\usepackage{algorithmic}
\usepackage{algorithm}

\title{\LARGE \bf
	Elbow Angle Guidance System Based on Surface Haptic Sensations \\Elicited by Lightweight Wearable Fabric Actuator
}

\author{Kenta Yokoe$^{1}$, Tadayoshi Aoyama$^{1}$, Yuki Funabora$^{2}$, Masaru Takeuchi$^{1}$ and Yasuhisa Hasegawa$^{1}$\\
\thanks{This work was supported by JST [Moonshot R\&D] Grant Number JPMJMS2214-08, JSPS KAKENHI Grant Numbers JP22H03630 and JP23KJ1116, and the Program for Promoting the Enhancement of Research Universities.}
\thanks{The protocol of this study was approved by the Ethics Committee of the Graduate School of Engineering, Nagoya University (Approval Number 23-11).}
\thanks{$^{1}$K. Yokoe, T. Aoyama, M. Takeuchi, and Y. Hasegawa are with the Department of Micro-Nano Mechanical Science and Engineering, Nagoya University, Nagoya, Aichi 464-8603, Japan.
	{\tt\small yokoe@robo.mein.nagoya-u.ac.jp; tadayoshi.aoyama@mae.nagoya-u.ac.jp; masaru.takeuchi@mae.nagoya-u.ac.jp; hasegawa@mein.nagoya-u.ac.jp}}%
\thanks{$^{2}$Y. Funabora is with the Department of Information and Communication Engineering, Nagoya University, Nagoya, Aichi 464-8603, Japan.
	 {\tt\small funabora@nagoya-u.jp}}%
\thanks{This is the accepted version of the paper published in Proc. 2024 IEEE International Conference on Advanced Intelligent Mechatronics (AIM), Boston, MA, USA, pp.~1447--1454, 2024. DOI: 10.1109/AIM55361.2024.10637238.
	\copyright~2024 IEEE. Personal use of this material is permitted. Permission from IEEE must be obtained for all other uses, in any current or future media, including reprinting/republishing this material for advertising or promotional purposes, creating new collective works, for resale or redistribution to servers or lists, or reuse of any copyrighted component of this work in other works.}
}

\begin{document}

	\maketitle
	\thispagestyle{empty}
	\pagestyle{empty}

	\begin{abstract}
		The demand for wearable haptic devices has rapidly increased for various applications. 
However, many haptic devices interfere with the wearer's activities and movements. 
In addition, several haptic devices fail to elicit intuitive haptic sensations by adjusting to the natural posture of the wearer.
To address these issues, we propose an elbow angle guidance system using a lightweight wearable fabric actuator. 
The proposed actuator is made of fabric and has two McKibben-type artificial muscles attached to it, rendering it extremely lightweight and facilitating the delivery of surface haptic sensations to intuitively induce elbow extension and flexion. 
The surface haptic sensation elicited by the fabric actuator is adjusted to natural body movements without interfering with the wearer's movements.
Moreover, the proposed system measures and guides the elbow angle by changing the intensity of the surface haptic sensation delivered to users in real time.
The accuracy of the proposed system is demonstrated through experiments involving human participants.
	\end{abstract}

	\section{Introduction}
The demand for haptic devices for rehabilitation and teleoperation based on virtual reality has notably increased \cite{wang2020Multimodal,gunther2019PneumAct,majidifardvatan2021review}. 
In particular, many wearable haptic devices for parts of the upper limbs, such as the elbow, have been developed \cite{bardi2022Upper}. 
Several requirements must be satisfied for such haptic devices. For instance, the haptic devices should not fatigue the wearer. In addition, the elicited haptic sensations should not interfere with the user's motion, and the sensations should be intuitive for the user to interpret.
However, few haptic devices satisfy such requirements \cite{pacchierotti2017Wearable,stephens-fripp2018Reviewa}. 

We propose an intuitive elbow angle guidance system using a lightweight haptic device that does not interfere with the user's motion.
We use a McKibben-muscle-actuated fabric as the haptic device to guide the user's elbow angle. The fabric actuator establishes a lightweight haptic device that combines the fabrics with McKibben artificial muscles, which are driven by pneumatic pressure actuators~\cite{funabora2017Prototype,funabora2018Flexible}. These actuators can deliver surface haptic sensations that do not interfere with natural body movements~\cite{peng2023FunabotSuit}. 
\begin{figure}[!t]
	\centering
	\includegraphics[keepaspectratio=true,width=.7\linewidth]{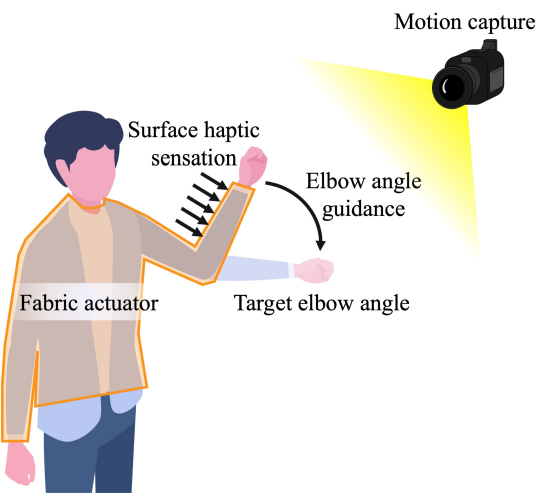}
	\caption{Schematic of proposed elbow angle guidance system.}
	\label{fig:concept}
\end{figure}
Fig.~\ref{fig:concept} shows a schematic of the proposed system. 
The system measures the user's elbow angle using an optical motion capture system and delivers surface haptic sensations from the fabric actuator to the upper limb based on the difference between the target and current elbow angles.
The elbow angle is measured sequentially, and the intensity of the surface haptic sensation is adjusted in real time.
The accuracy of elbow angle guidance in the proposed system was evaluated through an experiment involving human participants.

\section{Related Work}
Numerous exoskeleton designs have been proposed for elbow angle guidance.
Chen et al. proposed a two-degree-of-freedom (DoF) tethered exoskeleton using two identical serial elastic actuators for rehabilitation \cite{chen2019Elbow}. This exoskeleton allows the wearer to independently control the torque of elbow flexion/extension and forearm supination/pronation. 
Trigili et al. presented a four-DoF shoulder--elbow exoskeleton, called NeuroExos shoulder--elbow module, for upper-limb neurorehabilitation and spasticity treatment \cite{trigili2019Design}.
Dindorf et al. proposed a wearable orthosis for the elbow joint using a bi-muscular pneumatic servo drive \cite{dindorf2020Using}. The wearable orthosis is controlled through measured bioelectric signals. 
Liu et al. developed an exoskeleton for upper-limb power assistance \cite{liu2021wearable}. This exoskeleton assists in the flexion/extension of the elbow and shoulder. 
Zahedi et al. presented a wearable elbow exoskeleton for tremor suppression \cite{zahedi2021wearable}. This system suppresses tremors in the elbow joint using magnetorheological fluids. 
Park et al. introduced ElbowsideWINDER, an exoskeleton that aids elbow flexion and extension during occupational tasks \cite{park2023ElbowsideWINDER}. 
However, the abovementioned exoskeletons often restrict human movement because these exoskeletons are rigid devices. 

Many soft haptic devices have been designed to be attached to the elbow and avoid limiting the range of motion of the wearer.
Copani et al. designed a wearable exoskeleton for medical elbow rehabilitation using shape-memory alloy actuators \cite{copaci2017New}.
Although the exoskeleton is lightweight, real-time control is difficult because of the slow heating and cooling required to move the shape memory alloy. Therefore, no experiments with human participants have been conducted to evaluate the system.
Wei et al. developed a Bowden-cable-driven upper-limb soft exoskeleton \cite{wei2018Design}. It provides support during elbow flexion through three Bowden cables placed at the elbow. However, it can only support one DoF and cannot guide the elbow angle.
Wu et al. proposed a soft wearable exoskeleton for elbow assistance \cite{wu2019Neuralnetworkenhanced}. It includes surface electromyography sensors, inertial measurement units, force sensors, and a motor encoder to estimate the joint torque at the elbow and provide haptic feedback. 
Proietti et al. developed a textile-based multi-joint soft wearable robot to assist upper-limb motion \cite{oneill2020Inflatable,proietti2021Sensing}. This wearable robot includes pneumatic textile-based actuators and inertial measurement units and assists in shoulder elevation and elbow extension. 
Nassour et al. proposed an exoskeleton that combines a soft human--machine interface and soft pneumatic actuation to assist the elbow during load holding and carrying \cite{nassour2021Soft}. Experimental results with human participants confirmed that the muscle activity, metabolic rate, and fatigue were significantly reduced when using this assistance.
Jeong et al. proposed a soft wearable robot composed of shape-memory-alloy artificial muscles to assist upper-limb motion \cite{jeong2022Soft}. The wearable robot assists with elbow flexion and forearm supination/pronation. 
Mucchiani et al. proposed a pneumatically actuated soft wearable device to assist shoulder adduction/abduction and elbow flexion/extension \cite{mucchiani2022Closedloopa}. Experiments using an engineered mannequin demonstrated that elbow joint movements above 45 deg were difficult to perform using this device.
Hinchet et al. proposed a thin haptic sleeve based on electrostatic clutches \cite{hinchet2022Glove}. This haptic sleeve allows users to extend their elbows.
Ramachandran et al. developed a wearable haptic sleeve that covers the wrist and elbow \cite{ramachandran2022ArmWrista}. This sleeve is composed of electro-adhesive clutches, restricts the motion of the elbow and wrist, and assists in teleoperation of a drone.
Many haptic devices with soft actuators can support elbow torque. However, controlling the elbow angle using soft haptics is difficult and has not yet been tested in humans.
Additionally, many soft devices are not intuitive and require practice to familiarize the users with haptic sensations.

A McKibben-muscle-actuated fabric is a pneumatically driven, soft, and lightweight device~\cite {funabora2017Prototype,funabora2018Flexible}. 
This device can move clothing by attaching McKibben-type artificial muscles~\cite{nakagawa2022Turninga}.
By applying this fabric as a haptic device, we have induced torso forward/backward flexion and left/right twisting with intuitive operation~\cite{peng2023FunabotSuit}. 
In previous studies, the range of haptic sensation was confined to the torso, and the pneumatic pressure values were fixed prior to the presentation of the haptic sensation.
In this paper, we propose an artificial muscle arrangement and algorithm for pneumatic pressure adjustment method to establish a wearable haptic device that can intuitively guide the elbow angle without interfering with the user movement. 
The proposed artificial muscle arrangement enables the user to perceive the surface haptic sensation at the elbow. Furthermore, the pneumatic pressure adjustment method allows for the adjustment of the pneumatic pressure based on the user's elbow angle and the target angle.

\section{Lightweight Wearable Fabric Actuator for Elbow Angle Guidance}
\label{sec:system}
\begin{figure}[!t]
	\centering
	\includegraphics[keepaspectratio=true,width=\linewidth]{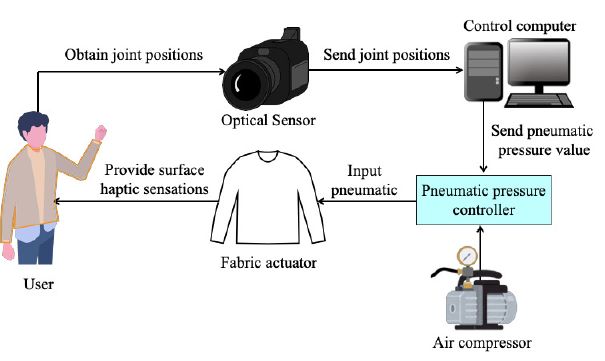}
	\caption{Configuration of elbow angle guidance system. }
	\label{fig:config}
\end{figure}
Fig.~\ref{fig:config} shows the configuration of the proposed elbow angle guidance system. 
The proposed system is composed of a laptop (Windows 10 Pro, 64 bits, Intel (R) Core (TM) i7-11800H CPU at 2.30 GHz, 16 GB RAM, NVIDIA GeForce RTX 3060 GPU), air compressor (TAKAGI ACP-39SLB), optical motion capture system (NaturalPoint OptiTrack), pneumatic pressure controller, and fabric actuator. The pneumatic pressure controller consists of programmable logic controllers (Industrial Shields 007001001100) and electric regulators (KOGANEI CRCB-0135W/0136W). The fabric actuator comprises a Velcro suit (NaturalPoint motion capture suit), two McKibben artificial muscles (EMM20~$\times$~800, s-muscle), and Velcro tape. The Velcro suit is the same as that used in~\cite{peng2023FunabotSuit}. Therefore, haptic sensations can be delivered to multiple body parts with a single device when combined with the artificial muscle arrangement used in the previous study.
The proposed fabric actuator uses Velcro tape to attach the artificial muscles, but they can be attached by any method.
The user wears the fabric actuator, and the motion capture system acquires the positions of the user's joints. 
The current elbow angle is calculated from the acquired joint positions, and pneumatic pressure is applied to the fabric actuator based on the target and current elbow angles. The fabric actuator provides surface haptic sensations to the user to guide the elbow angle.

\begin{figure}[!t]
	\centering
	\includegraphics[keepaspectratio=true,width=\linewidth]{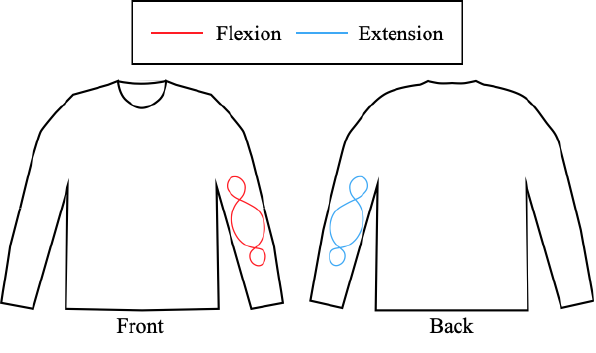}
	\caption{Artificial muscle arrangement of fabric actuator on Velcro suit.}
	\label{fig:muscle}
\end{figure}
\begin{figure}[!t]
	\centering
	\includegraphics[keepaspectratio=true,width=\linewidth]{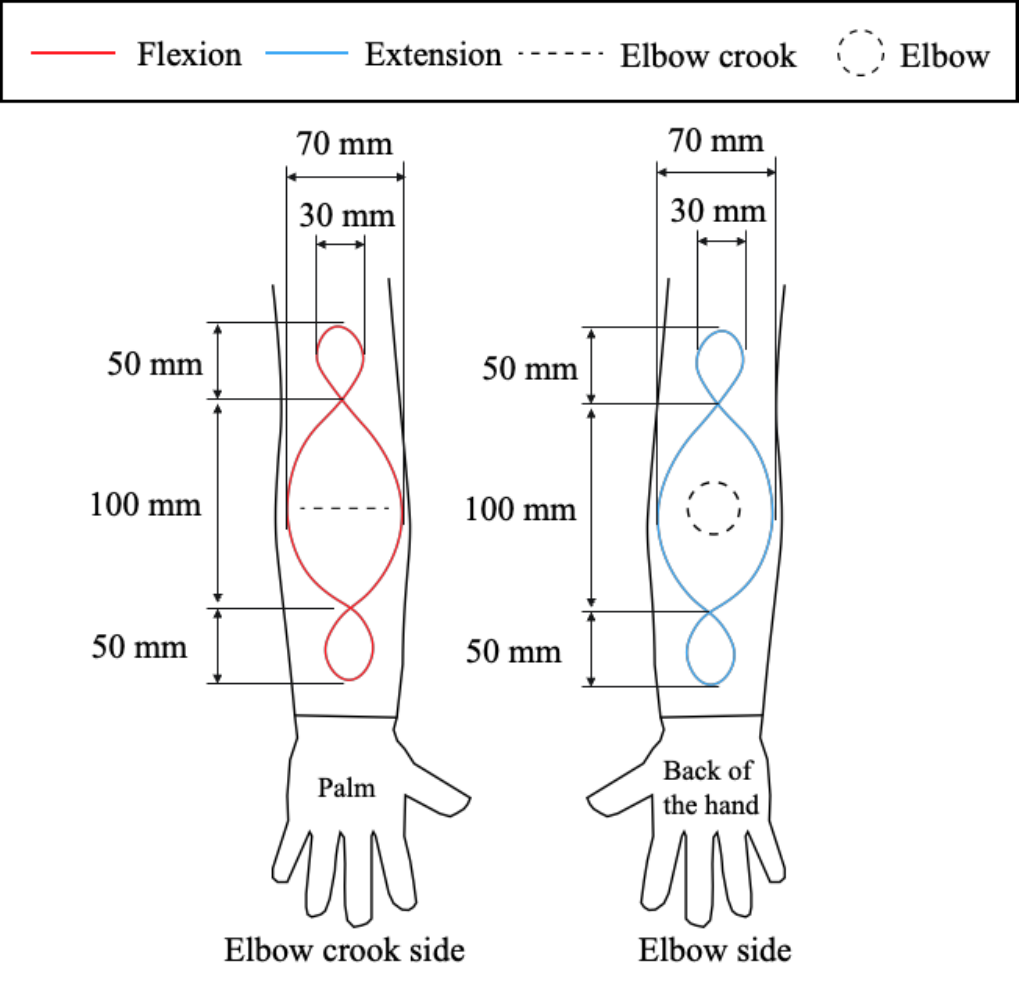}
	\caption{Schematic of artificial muscle arrangement of fabric actuator.}
	\label{fig:detail}
\end{figure}
Fig.~\ref{fig:muscle} shows the arrangement of the McKibben artificial muscles of the fabric actuator on the Velcro suit, and Fig.~\ref{fig:detail} shows a detailed schematic of the arrangement. 	The artificial muscle shown in red is placed linearly and symmetrically around the elbow crook on that side of the arm.
The artificial muscle shown in blue is placed linearly and symmetrically around the elbow on the other side of the arm.
The placements and lengths shown in Fig.~\ref{fig:detail} can be adjusted according to the arm dimensions.
Applying pneumatic pressure to each artificial muscle causes the muscle to move the fabric, thus delivering haptic sensations to the wearer.
When pneumatic pressure is applied to the artificial muscles shown in red, the fabric elicits a haptic sensation on the skin near the biceps brachii, brachialis, and brachioradialis, which are the muscles used for elbow flexion.
When pneumatic pressure is applied to the artificial muscles shown in blue, the fabric elicits a haptic sensation on the skin near the triceps brachii and anconeus, which are the muscles used for elbow extension.

\begin{figure}[!t]
	\centering
	\includegraphics[keepaspectratio=true,width=.9\linewidth]{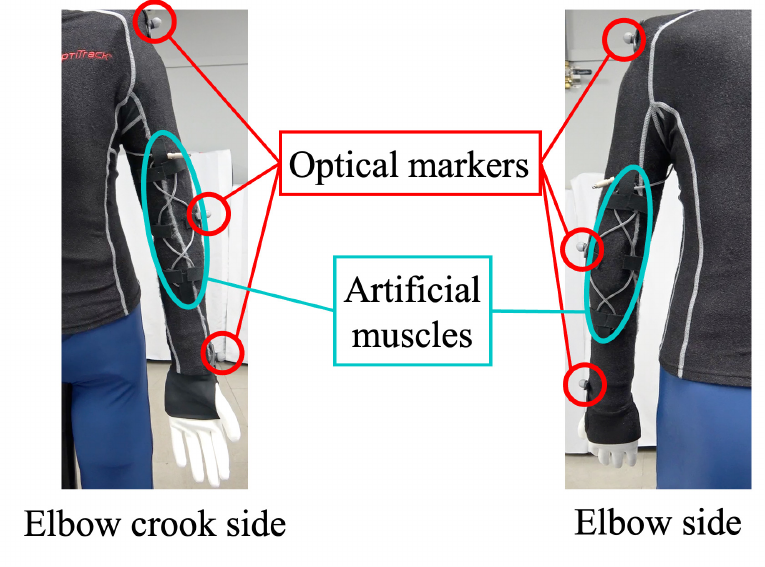}
	\caption{Photographs of fabric actuator for elbow angle guidance. Optical sensors (red circles) for elbow angle acquisition and artificial muscles (cyan circles) for eliciting haptic sensations are attached to the suit.}
	\label{fig:suit}
\end{figure}
\begin{figure}[!t]
	\centering
	\includegraphics[keepaspectratio=true,width=\linewidth]{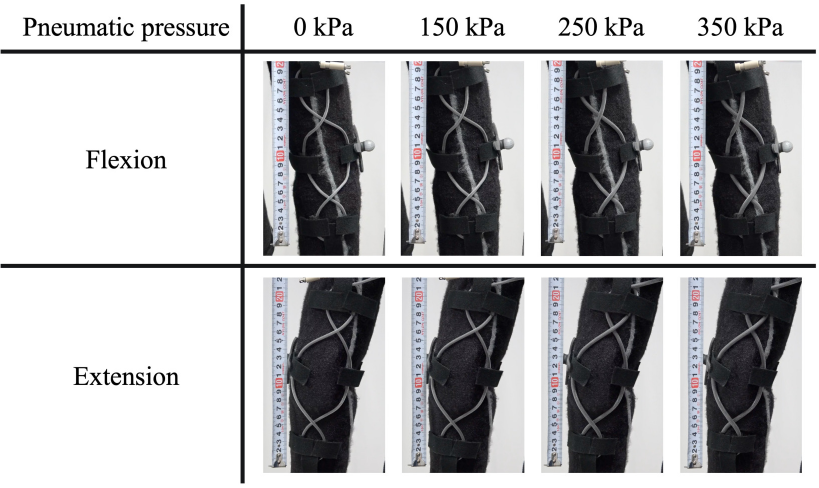}
	\caption{Deformations of artificial muscles and fabric according to pneumatic pressure for inducing elbow flexion and extension.}
	\label{fig:movements}
\end{figure}
Fig.~\ref{fig:suit} shows a fabric actuator worn on a real-care practice mannequin (Human Doll Corporation Yururi-kun). 
The mannequin has joints that move as those of a human.
Fig.~\ref{fig:movements} shows the movements of the fabric actuator worn on the mannequin when pneumatic pressure is applied. 
No notable difference is observed between no applied pneumatic pressure and a pressure of 150 kPa in the artificial muscle actuation.
As the applied pneumatic pressure increases from 150 kPa to 350 kPa, the diameter of the artificial muscle increases, thus contracting the muscle and moving fabric. Thus, the wearer likely perceives surface haptic sensations elicited by the fabric actuator.
Even under 350 kPa, the elbow of the mannequin cannot be flexed or extended by more than 1 deg.
Therefore, the proposed fabric actuator does not force elbow motion and is thus less likely to interfere with the wearer's movement.

\section{Pneumatic Pressure Adjustment Based on Elbow Angle}
For the proposed system, the relation between the elbow angle and intensity of the surface haptic sensation generated by the fabric actuator to guide the elbow should be determined.
The intensity of the surface haptic sensation is positively related to the magnitude of the pneumatic pressure applied to the fabric actuator. Thus, an equation that relates the input pneumatic pressure to the elbow angle should be obtained.

According to Weber–Fechner's law \cite{colman2009Dictionary}, human perception $E$ and stimulus intensity $I$ can be related as follows: 
\begin{align}\label{eq:fech}
	E=k\ln (\frac{I}{I^0}),
\end{align}
where $k$ is a constant and $I^0$ is the threshold intensity of the stimulus when human perception $E$ is zero.
The inverse function of Equation~\eqref{eq:fech} is expressed as 
\begin{align}\label{eq:rev}
	I=I^0e^\frac{E}{k}. 
\end{align}

\begin{figure}[!t]
	\centering
	\includegraphics[keepaspectratio=true,width=.85\linewidth]{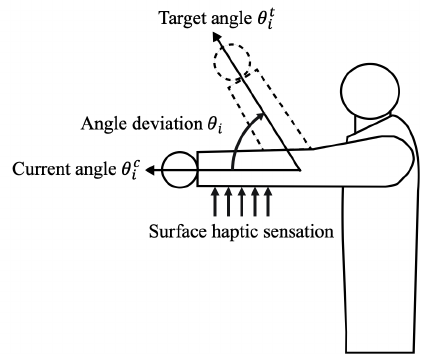}
	\caption{Schematic of elbow angle guidance (left view). The user's arm is initially extended.}
	\label{fig:alg}
\end{figure}
Fig.~\ref{fig:alg} shows a schematic of the elbow angle guidance using the fabric actuator.
Let $\theta_i^t$ $(i=1, 2)$ be the target angle of the elbow and $\theta_i^c$ $(i=1, 2)$ be the current angle of the elbow. We define $\theta_i$ $(i=1, 2)$ as 
\begin{align}\label{eq:theta}
	\theta_i=\theta_i^t-\theta_i^c \hskip 2 mm (i=1, 2), 
\end{align}
where $i$ equal to 1 or 2 indicates elbow flexion or extension, respectively.
When $i$ = 1, the direction of elbow flexion is considered positive, and when $i$ = 2, the direction of elbow extension is considered positive. 
Adapting Equation~\eqref{eq:rev}, the relation between the pneumatic pressure applied to the fabric actuator, $P_i$ $(i=1, 2)$, and elbow angle $\theta_i$ $(i=1, 2)$ can be expressed as 
\begin{align}\label{eq:air}
	P_i=f(\theta_i)=P^0_ie^{\frac{\theta_i}{k_i}} \hskip 2 mm (i=1, 2),
\end{align}
where $k_i$ $(i=1, 2)$ represents a constant value and $P^0_i$ $(i=1, 2)$ represents the threshold magnitude of the pneumatic pressure applied to the fabric actuator when the user does not perceive a surface haptic sensation. 
$P ^0_i$ $(i=1, 2)$ must be determined by using Equation~\eqref{eq:air} to control the pneumatic pressure applied to the fabric actuator.
Thus, the proposed system uses Algorithm~\ref{alg:fitting} based on the method of limits \cite{colman2009Dictionary} to determine $P ^0_i$ $(i=1, 2)$ by applying Equation~\eqref{eq:air}.
\begin{algorithm}[!t]
	\caption{Calculation of $P ^0_i$ $(i=1, 2)$ using Equation~\eqref{eq:air}.}
	\label{alg:fitting}
	\begin{algorithmic}[1]
		\renewcommand{\algorithmicrequire}{\textbf{Input:}}
		\REQUIRE  $i = 1, 2$
		\FOR  {$i=1$ to $2$}
		\FOR {$j=1$ to $P^{MAX}_i$}
		\STATE $P_{ij} \gets j$
		\STATE Apply pneumatic pressure $P_{ij}$ to artificial muscle
		\IF {User perceives surface haptic sensation \OR j = $P^{MAX}_i$}
		\STATE $P^1_i \gets P_{ij}$
		\STATE Exit current For statement 
		\ENDIF 
		\ENDFOR
		\FOR {$j=1$ to $P^{MAX}$}
		\STATE $P_{ij} \gets P^{MAX}_i - j$
		\STATE Apply pneumatic pressure $P_{ij}$ to artificial muscle
		\IF {User does not perceive surface haptic sensation \OR \\$j = P^{MAX}_i$}
		\STATE $P^2_i \gets P_{ij}$
		\STATE Exit current For statement 
		\ENDIF
		\ENDFOR
		\STATE $P_i^0 \gets \frac{P^1_i+P^2_i}{2}$
		\ENDFOR
	\end{algorithmic}
\end{algorithm}
In Algorithm~\ref{alg:fitting}, $P^{MAX}_i$ $(i = 1, 2)$ is the maximum pneumatic pressure applied to the fabric actuator. 
To determine $P^0_i$ $(i=1, 2)$, increasing pneumatic pressure is applied to the fabric actuator, starting from the minimum pressure. The proposed system records the pneumatic pressure when the user first perceives the surface haptic sensation as $P^1_i$ $(i=1, 2)$. 
Subsequently, decreasing pneumatic pressure is applied to the fabric actuator, starting from the maximum pressure. The proposed system records the pneumatic pressure when the user no longer perceives the surface haptic sensation as $P^2_i$ $(i=1, 2)$. 
In the proposed system, $P^0_i$ $(i=1, 2)$ is defined as the mean of $P^1_i$ $(i=1, 2)$ and $P^2_i$ $(i=1, 2)$.
These pressures are determined for surface haptic sensations guiding both the elbow extension and flexion to obtain the corresponding $P^0_i$ $(i=1, 2)$ for Equation~\eqref{eq:air}.

The proposed system defines $k_i$ $(i=1, 2)$ by using the maximum pneumatic pressure applied to the fabric actuator, $P^{MAX}_i$ $(i = 1, 2)$, and $\theta^{MAX}_i$ $(i = 1, 2)$, which is the maximum value of $\theta_i$ $(i=1, 2)$ in Equation~\eqref{eq:theta}, as follows: 
\begin{align}\label{eq:k}
	k_i=\frac{\theta^{MAX}_i}{\ln (\frac{P^{MAX}_i}{P^0_i})} \hskip 2 mm (i=1, 2).
\end{align}
By defining $k_i$ as in Equation~\eqref{eq:k}, the induced elbow angle is $\theta^{MAX}_i$ $(i = 1, 2)$ when pneumatic pressure $P^{MAX}_i$ $(i = 1, 2)$ is applied to the fabric actuator.

\section{Experiment to Evaluate Accuracy of Elbow Angle Guidance}
\subsection{Experimental Setup}
\begin{figure}[!t]
	\centering
	\includegraphics[keepaspectratio=true,width=.8\linewidth]{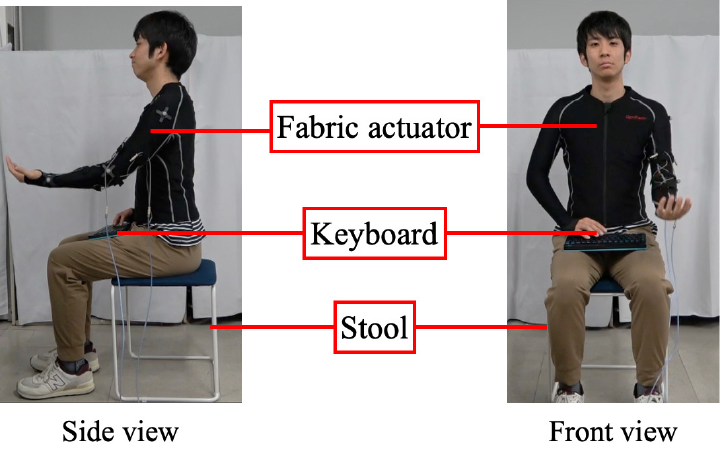}
	\caption{Participant viewed from side and front with elbow angle of 135 deg.}
	\label{fig:subject}
\end{figure}

We conducted an experiment with human participants to evaluate the accuracy of the elbow angle guidance system. Six participants were enrolled in this study. They were healthy and right-handed and aged between 20 and 25 years. The protocol of this experiment was approved by the Ethics Committee of the Graduate School of Engineering, Nagoya University (Approval Number 23-11). 
Fig.~\ref{fig:subject} illustrates the experimental setup with a participant.
\begin{figure}[!t]
	\centering
	\includegraphics[keepaspectratio=true,width=\linewidth]{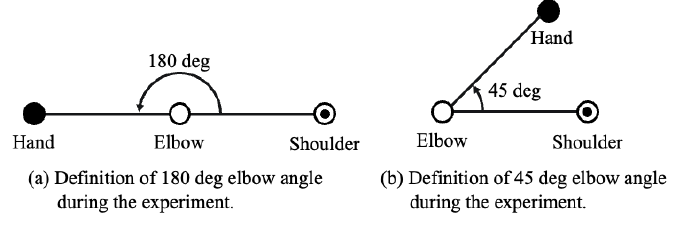}
	\caption{Definition of elbow angles for experiment. }
	\label{fig:difine}
\end{figure}
Every participant wore the proposed fabric actuator, sat on a stool, and placed a keyboard on their lap.

In the experiment, the proposed system guided the participant's left elbow angle by eliciting surface haptic sensations on the left arm. The participant moved the elbow following the haptic sensation and tapped the keyboard when they finished moving the elbow. The final elbow angle was recorded, and the error between the target and actual elbow angles was measured.
The elbow angle was defined as 180 deg when the elbow was extended and the shoulder, elbow, and wrist were aligned, and 0 deg when the shoulder and wrist were close, as illustrated in Fig.~\ref{fig:difine}. 
The target elbow angle was randomly set in 45–180 deg because it was difficult to reach an elbow angle of less than 45 deg.
Therefore, $\theta^{MAX}_i$ $(i = 1, 2)$ in Equation~\eqref{eq:k} was set to 135 deg.
In the experiment, the maximum pneumatic pressure applied to the fabric actuator, $P^{MAX}_i$ $(i = 1, 2)$, was 350 kPa.
Before the experiment, the participants were instructed to set their elbow angles to 45 deg, 90 deg, 135 deg, or 180 deg in a relaxed position. 
At each initial elbow angle, $P^0_i$ $(i=1, 2)$ in Equation~\eqref{eq:air} was determined by applying Algorithm~\ref{alg:fitting}.
The experiment was performed 10 times per participant at each initial elbow angle. The initial elbow angle was determined randomly.
No practice was performed to ensure that the participants were unfamiliar with the fabric actuator.
In addition, the participants were not given any information about the haptic sensation, including its purpose of guiding the elbow flexion or extension or its intensity indicating the intended elbow angle.
In addition, a random elbow angle response to the same target elbow angle as in the experiment was generated by the Mersenne Twister for comparison with random chance~\cite{matsumoto1998Mersenne}.
\subsection{Experimental Results}
\begin{table*}[!t]
	\centering
	\caption{$P^0_1$ in Equation~\eqref{eq:air} used for elbow flexion per participant and initial elbow angle. $^{a}$ and $^{b}$ indicate significant differences by Bonferroni tests. }
	\label{tab:flexion}
	\begin{tabular}{c|cccccc|c}
		Initial elbow angle & A     & B     & C     & D     & E     & F     & Mean \\ \hline
		45                  & 133.0 & 149.0 & 146.5 & 176.5 & 139.5 & 157.5 & 150.3   \\
		90                  & 150.0 & 129.0 & 143.5 & 153.0 & 161.5 & 159.0 & 149.3   \\
		135                 & 146.0 & 119.5 & 155.0 & 162.0 & 151.0 & 182.5 & 152.7   \\
		180                 & 145.0 & 105.0 & 190.5 & 156.5 & 134.5 & 189.0 & 153.4   \\ \hline
		Mean             & 143.5 & 125.6$^{a, b}$ & 158.9 & 162.0$^{a}$ & 146.6 & 172.0$^{b}$ & 151.4  
	\end{tabular}
\end{table*}
\begin{table*}[!t]
	\centering
	\caption{$P^0_2$ in Equation~\eqref{eq:air} used for elbow extension per participant and initial elbow angle. }
	\label{tab:extension}
	\begin{tabular}{c|cccccc|c}
		Initial elbow angle & A     & B     & C     & D     & E     & F     & Mean \\ \hline
		45                  & 130.5 & 87.5  & 144.5 & 140.0 & 96.5  & 184.5 & 130.6   \\
		90                  & 160.5 & 138.5 & 150.0 & 151.0 & 153.0 & 128.0 & 146.8   \\
		135                 & 147.0 & 126.5 & 91.5  & 167.0 & 146.0 & 177.5 & 142.6   \\
		180                 & 143.0 & 110.5 & 130.0 & 168.5 & 215.5 & 182.0 & 158.3   \\ \hline
		Mean             & 145.3 & 115.8 & 129.0 & 156.6 & 152.8 & 168.0 & 144.6  
	\end{tabular}
\end{table*}
\begin{table*}[!t]
	\centering
	\caption{Median, mean, and standard deviation of absolute error per participant. }
	\label{tab:median}
	\begin{tabular}{c|cccccc|c|c}
		Participant& A    & B    & C    & D    & E    & F      & All  &Random responses\\ \hline
		Median    & 20.9 & 18.5 & 28.6 & 24.0 & 22.3 & 29.0 & 22.8&39.1\\
		Mean    & 25.3 & 21.5 & 31.9 & 24.3 & 23.7 & 31.0 & 26.3&44.2	\\
		Standard deviation& 19.6 & 13.7 & 19.2 & 16.0 & 18.5 & 22.4& 18.8&31.6
	\end{tabular}
\end{table*}
\begin{figure}[!t]
	\centering
	\includegraphics[keepaspectratio=true,width=.9\linewidth]{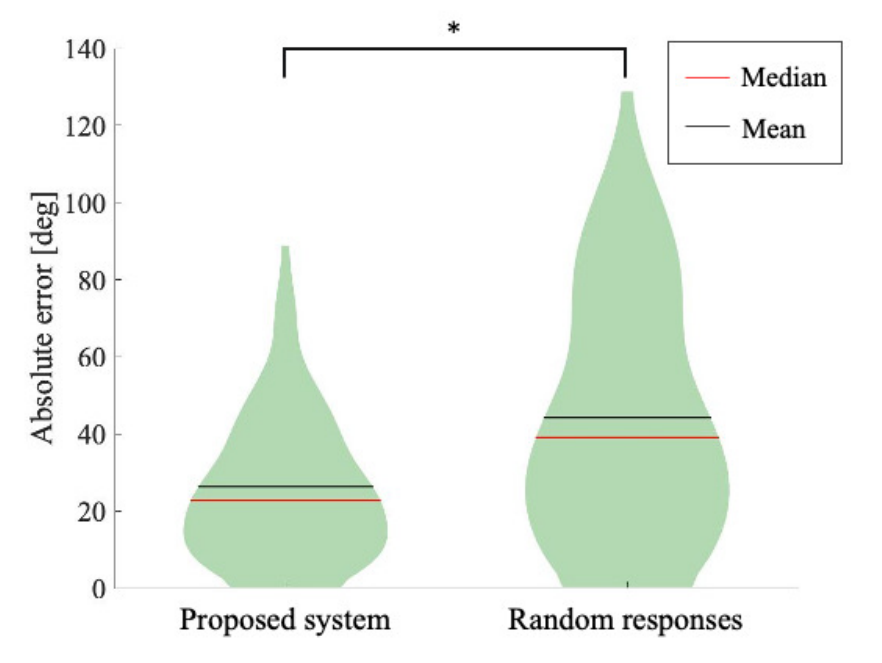}
	\caption{Violin plots of absolute error of elbow angle obtained from experiment (left) and random responses without using the proposed system for the same target angle as in the experiment (right). $*$ indicates a significant difference by the Wilcoxon rank sum test. }
	\label{fig:vsrandom}
\end{figure}
\begin{figure}[!t]
	\centering
	\includegraphics[keepaspectratio=true,width=.9\linewidth]{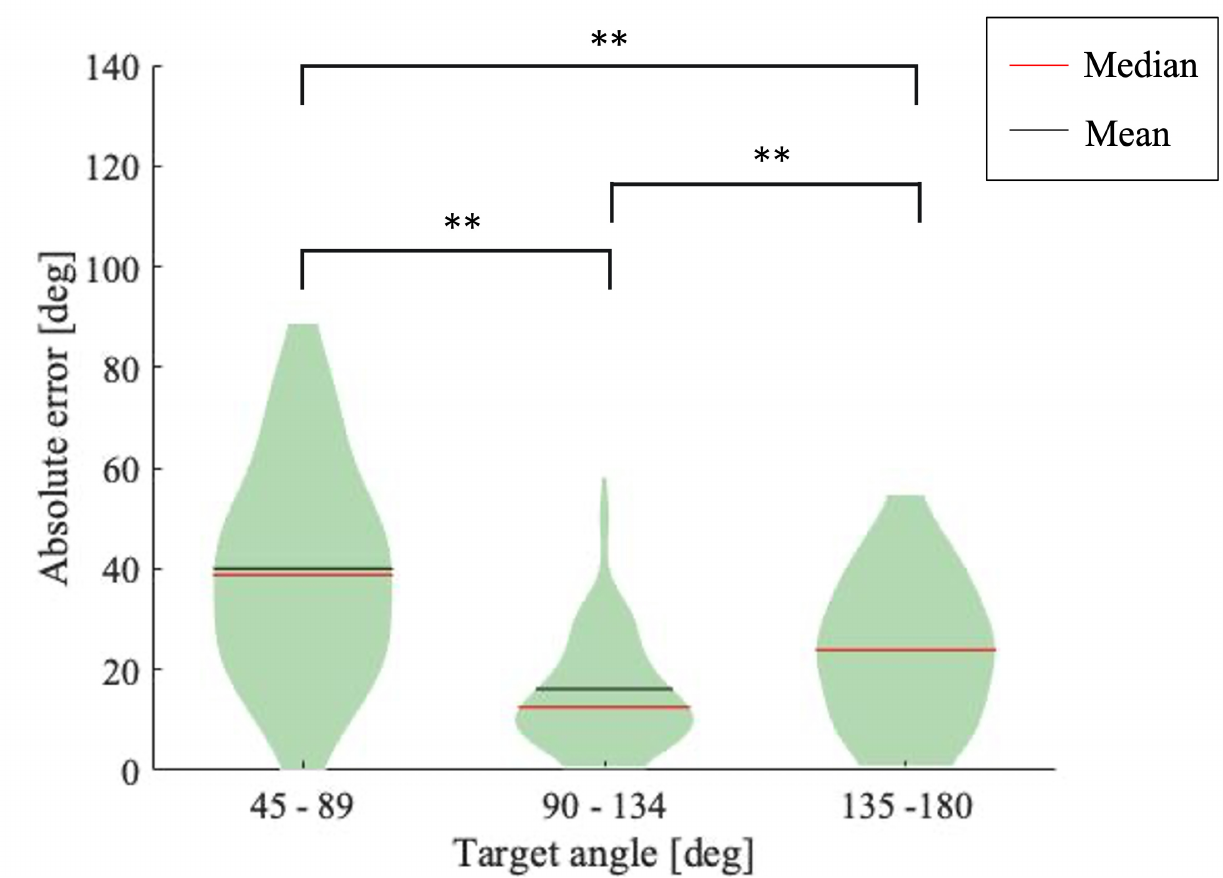}
	\caption{Violin plots of absolute errors of elbow angle according to target angles. $**$ indicates a significant difference by a Kruskal--Wallis test with Bonferroni correction. }
	\label{fig:pertarget}
\end{figure}
\begin{figure}[!t]
	\centering
	\includegraphics[keepaspectratio=true,width=.9\linewidth]{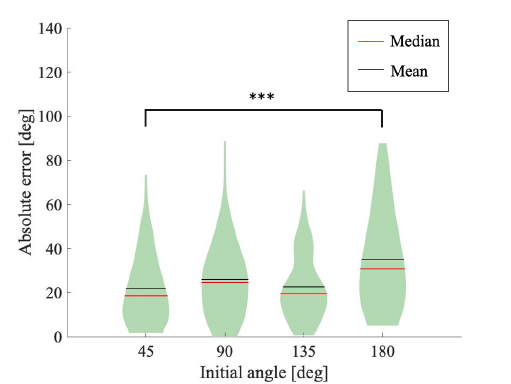}
	\caption{Violin plots of absolute errors of elbow angle according to initial angle. $***$ indicates a significant difference by a Kruskal--Wallis test with Bonferroni correction. }
	\label{fig:perinitial}
\end{figure}

Tables~\ref{tab:flexion} and \ref{tab:extension} list $P^0_1$ and $P^0_2$ (in kilopascals) in Equation~\eqref{eq:air}, respectively, per participant (from A to F) and initial elbow angle. 
Table~\ref{tab:median} lists the median, mean, and standard deviation (in degrees) of the absolute error between the target and actual elbow angles. The error values for all the participants in the experiment were smaller than those of the random responses. 
Fig.~\ref{fig:vsrandom} shows violin plots of the absolute error between the target and actual elbow angles obtained from the experiment using the proposed system and from simulating random response angles using Mersenne Twister~\cite{matsumoto1998Mersenne} for the same target elbow angle as in the experiment.
Figs.~\ref{fig:pertarget} and \ref{fig:perinitial} show the absolute errors per target and initial elbow angle, respectively. The target angles were divided into three ranges of 45 deg.

There were no significant differences in the mean values of $P^0_i$ $(i=1, 2)$ in Equation~\eqref{eq:air}, as indicated by the initial elbow angles for both flexion and extension (one-way ANOVA, $p=0.986$ for flexion and $p=0.492$ for extension).
Therefore, the initial elbow angle did not affect $P^0_i$ $(i=1, 2)$.
In addition, the paired t-test results showed no significant difference between $P^0_1$ in Equation~\eqref{eq:air} for elbow flexion and $P^0_2$ in Equation~\eqref{eq:air} for elbow extension (paired t-test, $p=0.292$).
There was no significant difference in sensitivity between the haptic sensations for flexion and extension.
The comparisons between participants showed no significant difference in $P^0_2$ for elbow extension (one-way ANOVA, $p=0.143$). However, there was a significant difference in $P^0_1$ for elbow flexion (one-way ANOVA, $p < 0.01$). Therefore, multiple comparisons were performed using the Bonferroni test.
The test revealed significant differences between participants B and D ($p=0.048$) and between participants B and F ($p=0.006$).
Hence, participant B perceived haptic sensations with less applied pneumatic pressure than participants D and F.

A Wilcoxon rank-sum test revealed a significant difference between the absolute elbow angle errors using the proposed system and that obtained from the random responses ($p < 0.001$). Hence, the proposed system can guide the elbow angle.
The target elbow angles were divided into three groups: 45--89 deg, 90--134 deg, and 135--180 deg for multiple comparisons, and there were significant differences between all the groups (Kruskal--Wallis test with Bonferroni correction, $p < 0.001$). The absolute angle error was significantly lower when the target elbow angle was in 90--134 deg.
Multiple comparisons were performed using the Kruskal--Wallis test with Bonferroni correction for the distribution of the absolute error with the initial elbow angle as a factor. A significant difference was observed between 45 deg and 180 deg ($p = 0.028$). Specifically, the absolute error when the initial elbow angle was 45 deg was significantly smaller than that when it was 180 deg. 
There were no significant differences in the distribution of absolute errors between the participants (Friedman test with Bonferroni correction, $p > 0.05$). 
The effect of the surface haptic sensations did not differ significantly across participants.

\section{Discussion} 
The proposed system must obtain $P^0_i$ $(i=1, 2)$ from Equation~\eqref{eq:air}, which is the threshold of pneumatic pressure applied to the fabric actuator when the user does not perceive a surface haptic sensation. 
The experimental results showed that the initial elbow angle and surface haptic sensation indicating either flexion or extension did not affect the value of $P^0_i$ $(i=1, 2)$. 
As McKibben-type artificial muscles moved the fabric, the fabric actuator delivered surface haptic sensations to the wearer. Because the magnitude of the pneumatic pressure at which the artificial muscles began moving the fabric did not change significantly at any elbow angle, $P^0_i$ $(i=1, 2)$ was likely independent of the initial elbow angle.
In the proposed fabric actuator, the artificial muscles that induced flexion and extension of the elbow had the same length and were arranged similarly. Therefore, the magnitudes of pneumatic pressure at which the artificial muscles began moving the fabric were similar. 
Thus, $P^0_i$ $(i=1, 2)$ was independent of whether the surface haptic sensation indicated flexion or extension. 

Significant differences were observed for $P^0_1$ between certain participants, suggesting that sensitivity to surface haptic sensations differ across persons. 
The mean of $P^0_i$ $(i=1, 2)$ for participant B was the smallest among the participants, suggesting that participant B was the most sensitive to surface haptic sensations. 
This sensitivity may explain why participant B showed the smallest mean, median, and standard deviation of the absolute error among the participants.
In the experiment, no training was conducted to familiarize the participants with the surface haptic sensations. More accurate elbow angle guidance may be achieved by improving the participants' sensitivity to surface haptic sensations through training.
The determination of sensitivity to surface haptic sensations should be investigated in future work.

When using the proposed system, the absolute elbow angle error of all the participants was smaller than that of the random responses. Furthermore, there was a significant difference between the proposed system results and random responses.
Therefore, the proposed fabric actuator could guide the elbow angle of the wearer.
In addition, the user perceived surface haptic sensations delivered by the fabric actuator and intuitively determined whether these sensations indicated flexion or extension. 
However, the proposed system could not move the mannequin elbow, suggesting that it could only induce elbow movement by tactile sensations without exerting a large force that interfered with elbow movement.

The absolute error was the smallest when for target angles of 90--134 deg in the experiment.
In this range, the elbow did not need to be moved as much as in the 45--89 deg and 135--180 deg ranges, possibly explaining the obtained lowest absolute error.
When the target angle was 135--180 deg, the elbow angle was guided in the direction of gravity, whereas when the target angle was 45--89 deg, the elbow was guided against gravity. 
The proposed haptic sensations might had failed to deliver sufficient guidance to overcome gravity.
Accordingly, there was a significant difference between the initial elbow angles of 45 deg and 180 deg.
Therefore, the effectiveness of elbow guidance might be reduced by applying a stronger pneumatic pressure when inducing motion against gravity.

The median absolute error of the elbow angle obtained using the proposed system was 22.8 deg, while that obtained from the random response was 39.1 deg. Therefore, the proposed system succeeded in guiding the elbow angle, but its performance can be further improved.
Equation~\eqref{eq:air} was used for pneumatic pressure control in the proposed system. It determines the magnitude of the pneumatic pressure change according to the guiding elbow angle by adjusting $k _i$, which is in turn determined using Equation~\eqref{eq:k}:
In the experiment, $\theta ^{MAX}_i$ in Equation~\eqref{eq:k} was set to 135 deg to create a one-to-one correspondence between the target elbow angle and pneumatic pressure. However, in this case, the maximum pneumatic pressure, $P^{MAX}_i$, was applied to the fabric actuator with only two patterns: 1) the initial and target elbow angles were 45 deg and 180 deg, respectively, and 2) the initial and target elbow angles were 180 deg and 45 deg, respectively.
Therefore, by setting $\theta ^{MAX}_i$ to a smaller value and applying $P^{MAX}_i$ to every elbow angle exceeding $\theta ^{MAX}_i$, the elbow angle can be guided using a stronger surface haptic sensation.
Although such stronger sensation may improve the performance of elbow angle guidance, it may also inhibit user movement. 
Therefore, the pneumatic pressure applied to the fabric actuator must be adjusted carefully.

All the participants in the experiment were young and healthy.
Therefore, they were likely more sensitive to haptic sensations and had stronger muscles than older adults and infirm people.
If the proposed system is used for the elderly or infirm, stronger haptic sensations should be delivered than those elicited in this study.
In addition, the system cannot provide haptic sensations that are sufficiently strong to induce elbow movement. Thus, users with weakened muscles may not properly adjust the elbow angle.

\section{Conclusion}
We propose an elbow angle guidance system based on surface haptic sensations elicited by a lightweight wearable fabric actuator. 
The proposed fabric actuator comprises a Velcro suit and two McKibben-type artificial muscles. The actuator delivers surface haptic sensations that intuitively induce elbow flexion or extension without interfering with the wearer's motion. 
The proposed system implements a pneumatic pressure-adjustment algorithm for surface haptic sensations to guide the elbow joint angle based on the elbow angle obtained by a motion capture system. 
An experiment was conducted to evaluate the accuracy of the proposed elbow angle guidance system. 
The results confirmed that the proposed system can guide the elbow angle without requiring practice to familiarize the users with haptic sensations. 

Future work will include expanding the body regions where motion guidance can be provided by fabric actuators and applying the proposed elbow guidance system to virtual reality applications and rehabilitation.

\bibliography{AIM}
\bibliographystyle{IEEEtran}

\vfill
\end{document}